# The chaotic behavior of the black hole system GRS 1915+105

R. Misra[1], K.P. Harikrishnan[2], B. Mukhopadhyay[3], G. Ambika[4] and A. K. Kembhavi[1]

## ABSTRACT

A modified non-linear time series analysis technique, which computes the correlation dimension $D_2$, is used to analyze the X-ray light curves of the black hole system GRS 1915+105 in all twelve temporal classes. For four of these temporal classes $D_2$ saturates to $\approx 4-5$ which indicates that the underlying dynamical mechanism is a low dimensional chaotic system. Of the other eight classes, three show stochastic behavior while five show deviation from randomness. The light curves for four classes which depict chaotic behavior have the smallest ratio of the expected Poisson noise to the variability ($< 0.05$) while those for the three classes which depict stochastic behavior is the highest ($> 0.2$). This suggests that the temporal behavior of the black hole system is governed by a low dimensional chaotic system, whose nature is detectable only when the Poisson fluctuations are much smaller than the variability.

*Subject headings:* accretion, accretion disks - black hole physics - X-rays: binaries - X-rays: individual (GRS 1915+105)

## 1. Introduction

Black hole X-ray binaries are variable on a wide range of timescales ranging from months to milli-seconds. A detailed analysis of their temporal variability is crucial to the understanding of the geometry and structure of these high energy sources. Such studies may eventually be used to test the relativistic nature of these sources and to understand the physics of the accretion process. The variability in different energy bands is generally quantified by computing the power spectrum which is the amplitude squared of the Fourier transform.

[1]Inter-University Center for Astronomy and Astrophysics, Post Bag 4, Ganeshkhind, Pune-411007, India; rmisra@iucaa.ernet.in

[2]Dept. of Physics, The Cochin College, Cochin-682002, India

[3]Astronomy Division, P.O.Box 3000,University of Oulu, FIN-90014, Finland

[4]Dept. of Physics, Maharajas College, Cochin-682011, India

The power spectra give information about the characteristic frequencies of the system which show up as either breaks or as near Gaussian peaks, i.e Quasi-Periodic Oscillation (QPO) in the spectra (e.g. Belloni *et al.* (2001); Tomsick & Kaaret (2001); Rodriguez *et al.* (2002)). The shape of the power spectra, combined with the observed frequency dependent time lags between different energy bands, have put constraints on the radiative mechanisms and geometry of emitting regions (e.g. Nowak *et al.* (1999); Misra (2000); Cui (1999); Poutanen & Fabian (1999); Chakrabarti & Manickam (2000); Nobili *et al.* (2001))

These results are based on the response of the system to temporal variations whose origin is not clear. Important insight into the origin can be obtained by the detection and quantification of the possible non-linear behavior of the fluctuations. For example, the presence of stochastic fluctuations would favor X-ray variations driven by variations of some external parameters (like the mass accretion rate), or the possibility that active flares occur randomly. On the other hand, if the fluctuations can be described as a deterministic chaotic system, then inner disk instability or coherent flaring activity models will be the likely origin. A quantitative description of the temporal behavior can also be compared with time dependent numerical simulations of the accretion process and will help examine the physical relevance of these simulations.

The non-Gaussian and non-zero skewness values of the temporal variation of the black hole system Cygnus X-1 suggested that the variations are non-linear in nature (Thiel *et al.* 2001; Timmer *et al.* 2000; Maccarone & Coppi 2002). More rigorous tests were applied to the AGN ArK 564 (Gliozzi *et al.* 2002) which also suggested non-linear behavior. Nonlinear time series analysis(NLTS) seems to be the most convenient tool to check if the origin of the variability is chaotic, stochastic or a mixture of the two and has been adopted in several disciplines to study complex systems (e.g. human brain, weather) and predict their immediate future (Schreiber 1999). This technique has also been used earlier to analyze X-ray data of astrophysical sources. Based on a NLTS analysis of EXOSAT data, Voges *et al.* (1987) claimed that the X-ray Pulsar Her X-1 was a low dimensional chaotic system. However, Norris & Matilsky (1989) pointed out problems with that analysis since the source has a strong periodicity and the data analyzed had low signal to noise ratio. Lehto *et al.* (1993) used the NLTS technique to analyze EXOSAT light curves of several AGN, and found that only one, NGC 4051, showed signs of low dimensional chaos. A similar analysis on the noise filtered *Tenma* satellite data of Cyg X-1, suggested that the source may be a low dimensional chaotic system with large intrinsic noise (Unno *et al.* 1990). These analysis were hampered by small number of data points ($\lesssim 1000$) in the light curve and/or noise. Hence, the reported detection of low dimension chaos was only possible by rather subjective comparison of the results of the data analysis with those from simulated data of chaotic systems with noise.

The Galactic micro quasar GRS 1915+105 is a highly variable black hole system. It shows a wide range of variability (Chen *et al.* 1997; Paul *et al.* 1997; Belloni *et al.* 1997a) which required Belloni *et al.* (2000) to classify the behavior in no less than twelve temporal classes. In this work, our motivation is to determine the temporal property of this source by using a modified nonlinear time series analysis for each of these twelve classes. The different kinds of variability and its brightness ( the average RXTE PCA count rate ranges from $5000 - 32000$ counts/s) makes this source an ideal one to detect chaotic behavior.

In the next section we describe the technique used to determine the Correlation dimension. The results of the analysis are presented in §3, while in §4 the work is summarized and discussed.

## 2. The Non-Linear time series analysis

The algorithm normally employed in this analysis (Grassberger & Procaccia 1983) aims at creating an artificial or pseudo space of dimension $M$ with delay vectors constructed by splitting a scalar time series s(t) with delay time $\tau$ as

$$\vec{x}(t) = [s(t), s(t+\tau), ....., s(t+(M-1)\tau)] \quad (1)$$

The correlation sum or the correlation function is the average number of data points within a distance R from a data point,

$$C_M(R) \equiv \lim_{N\to\infty} \frac{1}{N(N-1)} \sum_{i}^{N} \sum_{j,j\neq i}^{N} \mathrm{H}(R - |\vec{x}_i - \vec{x}_j|) \quad (2)$$

where $\vec{x}_j$ is the position vector of a point belonging to the attractor in the M-dimensional space, $N$ is the number of reconstructed vectors and H is the Heaviside step function. The fractional dimension $D_2(M)$ is defined as

$$D_2 \equiv \lim_{R\to 0} d\mathrm{log}C_M(R)/d\mathrm{log}(R) \quad (3)$$

and is essentially the scaling index of $C_M(R)$ variation with $R$. $D_2(M)$ can be used to differentiate between different temporal behavior since for an uncorrelated stochastic system, $D_2 \approx M$ while for a chaotic system, $D_2(M) \approx$ constant for $M$ greater than a certain dimension $M_{max}$.

For a finite duration light curve, there are two complications that hinder the successful computation of $D_2(M)$. First, for small values of $R$, $C_M(R)$ is of order unity and the result there would be dominated by Poisson noise. Second, for large values of $R$, $C_M(R)$ will saturate to the total number of data points. Usually, these two effects are avoided in the log$C_M(R)$ versus log$R$ plot and the slope $D_2$ is obtained from the linear part of the curve. However, such an exercise is subjective especially for high dimensions. Here, we use a numerical scheme to compute $D_2$, which takes into account the above effects and at the same time optimizes the maximum use of the available data. The details of the method and several tests of its validity will be presented elsewhere ( Misra *et al. in preparation*). Briefly, the technique involves converting the original light curve to a uniform deviate, and to redefine the correlation function $C_M(R)$ as the average number of data points within a M-cube (instead of a M-sphere) of length $R$ around a data point. Only those M-cubes are considered which are within the embedding space, ensuring that there are no edge effects due to limited data points. This imposes a maximum value of $R < R_{max}$ for which $C_M(R)$ can be computed. To avoid the Poisson noise dominated region, only results from $R$ greater than a $R_{min}$ are taken into consideration such that the average $C(R_{min}) > 1$ where the Poisson noise would approximately be $1/\sqrt{N_c}$. Typically $C_M(R)$ is computed for ten different values of $R$ between $R_{min}$ and $R_{max}$ and the logarithmic slope for each point is computed and the average is taken to be $D_2(M)$. The error on $D_2(M)$ is estimated to be the mean standard deviation around this average. It should be noted that there often exists a critical $M_{cr}$ for which $R_{max} \approx R_{min}$ and no significant result can then be obtained for $M > M_{cr}$. Figure 1 (a) shows the $D_2(M)$ curve for a time series generated from random numbers and for the well known analytical low dimensional chaotic system, the Lorenz system. The total number of data points used to generate both curves is 30000 and the number of random centers used is $N_c = 2000$. As expected the $D_2$ plot for the random data is consistent with the $D_2 = M$ curve, while the plot for the Lorenz system shows significant deviation and saturates at $M \approx 3$ to a $D_2 \approx 2$, which is close to the known value of 2.04. The random data and the low dimensional chaotic system can clearly be distinguished in this scheme.

## 3. Results

The temporal property of GRS 1915+105 have been classified into twelve different classes by Belloni *et al.* (2000), who also present the observational dates and identification number of the RXTE data they had used to make the classification. Here, we have chosen a representative data for each class and extracted a few continuous data streams ($\approx$ 3000 sec long) from it. The observational IDs of the data used in this work are tabulated in Table 1. The light curves were generated with a resolution of 0.1 seconds resulting in $\approx$ 30000 data points for each of them and $\approx$ 1500 counts per bin. Light curves with finer time resolution are Poisson noise dominated, while larger binning gives too few data points.

In general, $D_2(M)$ is proportional to $\tau$ when $\tau$ is small and saturates (i.e. it is nearly invariant) for $\tau$ greater than a critical value and it is this saturated value which is the correct estimate of $D_2(M)$. As an example, the $D_2(M)$ curves for different values of $\tau$ are plotted in Figure 1 (b), where it can be seen that the curve is similar within error bars for $\tau = 15$, 25 and 100 sec. For all the data analyzed here the critical $\tau < 5-20$ sec, and hence the saturated curve (typically for $\tau \approx 50$ sec) is considered. It has been verified that the $D_2(M)$ curves for two separate light curves for the same class, are similar to within the error bars. This shows that as expected the temporal behavior of the system is more or less stationary for the same class. Hence such curves can be averaged to obtain a statistically more significant result.

Figure 2 shows the $D_2(M)$ curves for seven temporal classes. For four classes ($\lambda$, $\kappa$, $\beta$ and $\mu$) the curves show clear deviation from random behavior. For $\lambda$ and $\kappa$ there is saturation of $D_2 \approx 5$ for $M > 8$. For $\beta$ and $\mu$, the increase in $D_2$ is less than one when $M$ increases from 8 to 15. Thus these classes can be classified unambiguously as chaotic systems with correlation dimension less than 5 while the the behavior of the class $\phi$ is identical to a stochastic light curve. The classes $\alpha$ and $\rho$ show some deviation from stochastic behavior and hence this behavior, which is also seen in the classes $\theta$, $\nu$ and $\delta$, is named "non stochastic" in this work. As discussed below, these classes may be inferred to be low dimensional chaotic systems based on comparison with results from simulated data of chaotic systems with additional noise. Similar comparisons were made to infer the chaotic behavior of Cyg X-1 (Unno *et al.* 1990) and NGC 4051 (Lehto *et al.* 1993). We show in the last column of Table 1 the classification of all the twelve classes into one of these three categories, namely chaotic, non-stochastic and stochastic. We have listed in Table 1 the average counts $< S >$, the root mean square variation, the expected Poisson noise $< PN > \equiv \sqrt{< S >}$, and the ratio of the expected Poisson noise to the actual RMS value. It can be seen that there is a strong correlation between the inferred behavior of the system and the ratio of the expected Poisson noise to the rms values. This indicates that Poisson noise is affecting the analysis. To estimate the effect of Poisson noise, we consider the Lorenz system points $S_L(t)$ and rescale it by $S_{LR} = AS_L(t) + B$. A light curve is then simulated using $S_{LR}$ from the corresponding Poisson noise distributions. The constants $A$ and $B$ were chosen such that the simulated light curve had the same average count and rms variation as the two extreme cases for the GRS1915+105 data for $\beta$ and $\gamma$ classes. The results of the non linear time series analysis are shown in Figure 3, where it can be seen that even for the $\beta$ like case where the ratio of the expected Poisson noise to rms variation is only 4%, the $D_2$ versus $M$ curve saturates at a higher value than that of the original no noise data points. This implies that the correlation dimension of $\approx 4$ inferred from the analysis of the classes showing chaotic behavior (Figure 2) is an overestimation due to the inherent Poisson noise in the data. For larger Poisson noise fractions, the curve no longer saturates and becomes qualitatively similar to that obtained for the non-stochastic case.

## 4. Discussion

The saturation of the correlation dimension $D_2 \approx 4-5$ for four of the temporal classes clearly indicates that the underlying dynamic mechanism that governs the variability of the black hole system is a low dimensional chaotic one. As indicated by simulations of the Lorenz system with noise, the effect of Poisson noise in the data is to increase the $D_2$ values. Hence the real dimension of the system is probably smaller than $D_2 \approx 4-5$ that is obtained here. In fact it is possible that the the temporal behavior of the black hole system is always governed by a low dimensional chaotic system, but is undetectable when Poisson noise affects the analysis.

Alternatively, there may be a stochastic component to the variability which dominates for certain temporal classes. The two scenarios may be distinguished and better quantitative estimates of the correlation dimension may be obtained by either appropriate noise filtering of the data and/or appropriate averaging of the different light curves. Much longer ($\approx 30000$ sec long) continuous data streams sampled at 1 second resolution, would decrease Poisson noise and hence provide better quantitative measure of $D_2$. However, such long data streams are presently not available and merging non-continuous light curves, will require sophisticated gap filling techniques which might give rise to spurious results.

The variability of GRS 1915+105 can be interpreted as being transitions between three spectral states (Belloni *et al.* 2000), one of which (the so called soft state) is a long term canonical state observed in other black hole systems like Cygnus X-1 which do not show such high amplitude variability. It is attractive to identify these spectral states as fixed points which for GRS 1915+105 become unstable giving rise to the observed chaotic behavior which may also account for the ring like movement of the system in color-color space (Vilhu & Nevalainen 1998). The above hypothesis may be verified by future characterization of the chaos in GRS 1915+105. Note that GRS15+105 spends most of it's time in the $\chi$ class whose variability is similar to that observed in other black hole systems like Cygnus X-1. However, as shown in this work, Poisson noise effects the analysis for the $\chi$ class and the $D_2(M)$ values reflect stochastic behavior. This may be the reason why earlier different non-linear analysis of Cygnus X-1 data, while showing non-linearity (Timmer *et al.* 2000; Thiel *et al.* 2001) did not conclusively reveal chaotic behavior.

The identification of the temporal behavior of the black hole system as a chaotic one, has opened a new window toward the understanding of the origin and nature of their variability.

The present analysis can be extended to characterize the chaotic behavior. Using the minimum required phase space dimension, the data can be projected into different 2 dimensional planes, which will reveal the structure of the attractor and help to identify any possible centers of instability in the system. Further, dynamical invariants like the full Lyapunov spectrum, multi-fractal dimensions etc. can be also be computed. Recently, Winters *et al.* (2003) have studied and quantified the chaotic flow in magneto-hydrodynamic simulations of the mass accretion processes that is believed to be happening in black hole systems. The measured chaos parameters like the largest Lyapunov exponent for such simulations can be compared with that obtained from the light curve of black hole systems to validate such simulations and enhance our understanding of these systems. Note that such analysis can practically be applied only after the identification of the minimum phase space dimension which in turn usually requires the computation of $D_2(M)$.

GA and KPH acknowledge the hospitality and the facilities in IUCAA. BM thanks the Academy of Finland grant 80750 for support.

## REFERENCES

Belloni, T., Mendez, M., King, A. R., van der Klis, M., & van Paradijs, J., 1997a, ApJ, 479, L145.

Belloni, T., Klein-Wolt, M., Mendez, M., van der Klis, M., & van Paradijs, J., 2000, A&A, 355, 271.

Belloni, T., Mendez, M., & Sanchez-Fernandez, C., 2001, A&A, 372, 551.

Chakrabarti, S. K., & Manickam, S. G., 2000, ApJ, 531, L41.

Chen, X., Swank, J. H., & Taam, R. E., 1997, ApJ, 477, L41.

Cui, W., 1999, ApJ, 524, 59.

Gliozzi, M., *et al.*, 2002, A&A, 391, 875.

Grassberger, P. & Procaccia, I., 1983, *Physica D*, 9, 189.

Lehto, H. J., Czerny, B., & McHardy, I. M., 1993, MNRAS, 261, 125.

Maccarone, T. J., & Coppi, P. S., 2002, MNRAS, 336, 817.

Misra, R., 2000, ApJ, 529, L95.

Nobili, L., Belloni, T., Turolla, R., & Zampieri, L., 2001, Ap&SS, 276 , 217.

Norris, J. P., & Matilsky, T. A., 1989, ApJ, 346, 912.

Nowak, M. A., Vaughan, B. A., Wilms, J., Dove, J. B. & Begelman, M. C., 1999, ApJ, 510, 874.

Paul, B., *et al.* , 1997, A&A, 320, L37.

Poutanen, J., & Fabian, A. C., 1999, MNRAS, 306, L31.

Rodriguez, J., Durouchoux, P., Mirabel, I. F., Ueda, Y., Tagger, M., & Yamaoka, K., 2002, A&A, 386, 271.

Schreiber, T., 1999, Phys. Rep., 308, 1.

Thiel, M. *et al.* , 2001, Ap&SS, 276, 187.

Timmer, J., *et al.* , 2000, Phys. Rev. E, 61, 1342.

Tomsick, J. A., & Kaaret, P., 2001, ApJ, 548, 401

Unno, W., *et al.* , 1990, PASJ, 42, 269.

Vilhu, O., & Nevalainen, J., 1998, ApJ, 508, L85.

Voges, W., Atmanspacher, H., & Scheingraber, H., 1987, ApJ, 320, 794.

Winters, W. F., Balbus, S. A. & Hawley, J. F., 2003, MNRAS, 340, 519.

Table 1. Observation Table

| Obs. I.D. | Class | $<S>$ | rms | $<PN>$ | $<PN>$/rms | Behavior |
|---|---|---|---|---|---|---|
| 10408-01-10-00 | $\beta$ | 1917 | 1016 | 43.8 | 0.04 | C |
| 20402-01-37-01 | $\lambda$ | 1493 | 1015 | 38.6 | 0.04 | C |
| 20402-01-33-00 | $\kappa$ | 1311 | 800 | 36.2 | 0.04 | C |
| 10408-01-08-00 | $\mu$ | 3026 | 999 | 55 | 0.06 | C |
| 20402-01-45-02 | $\theta$ | 1740 | 678 | 41.7 | 0.06 | NS |
| 10408-01-40-00 | $\nu$ | 1360 | 462 | 36.9 | 0.08 | NS |
| 20402-01-03-00 | $\rho$ | 1258 | 440 | 35.5 | 0.08 | NS |
| 20187-02-01-00 | $\alpha$ | 582 | 244 | 24.1 | 0.10 | NS |
| 10408-01-17-00 | $\delta$ | 1397 | 377 | 37.4 | 0.10 | NS |
| 20402-01-56-00 | $\gamma$ | 1848 | 185 | 43.0 | 0.23 | S |
| 10408-01-22-00 | $\chi$ | 981 | 118 | 31.3 | 0.27 | S |
| 10408-01-12-00 | $\phi$ | 1073 | 118 | 32.7 | 0.28 | S |

Note
which
the cla
light cu
5: The
the exp
behavi
behavi
slightly

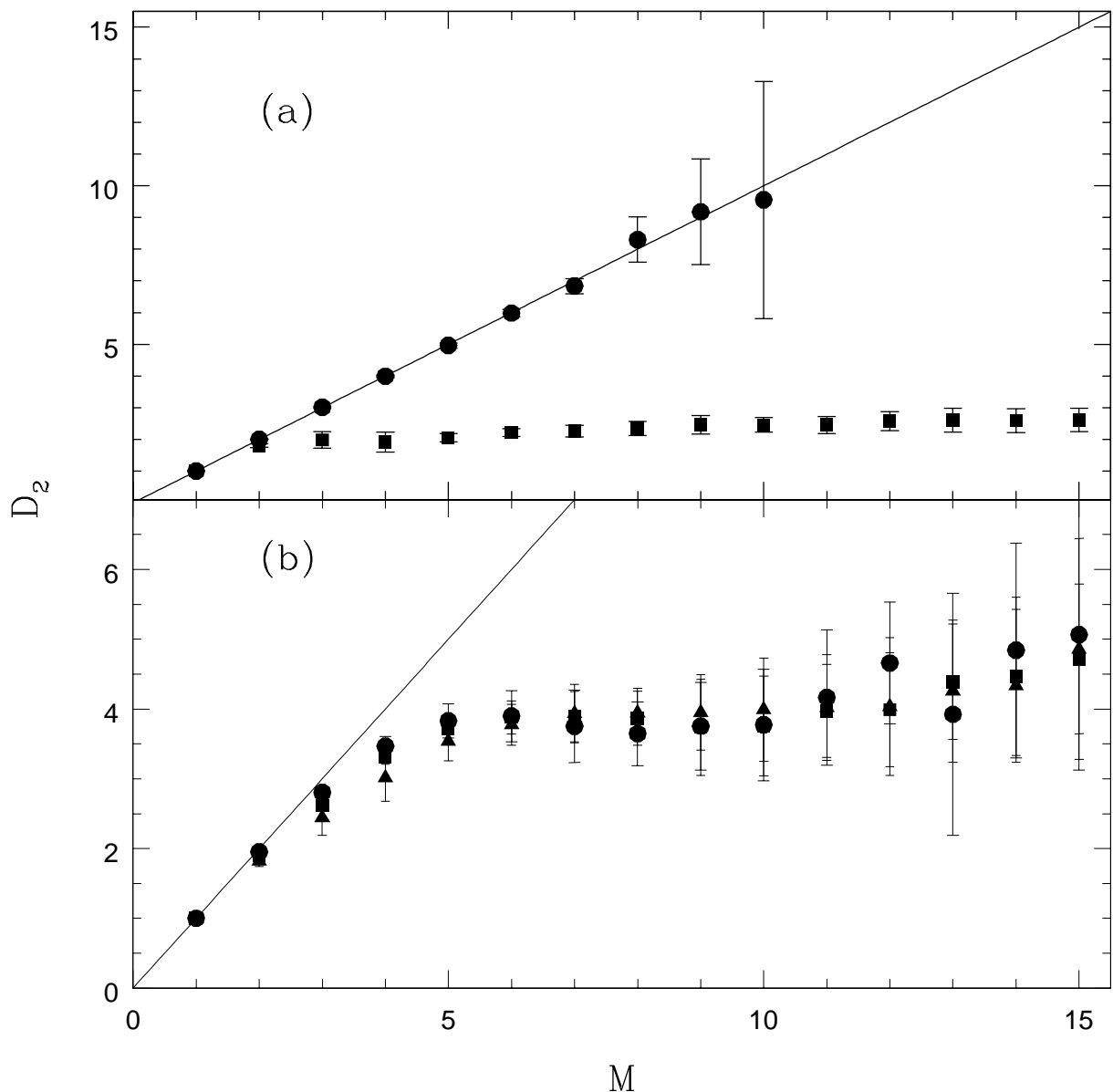

Fig. 1.— (a) The $D_2$ versus $M$ curve for random points (circles) and for Lorenz system (squares). For both the curves the number of points used are 30000 and the number of centers used in the computation is 2000. The straight line is the $D_2 = M$ case which is the expected result for random variation. (b) The $D_2$ versus $M$ curves for GRS1915+105 data obtained during the class $\kappa$ for three different values of the delay time $\tau = 15$ s (triangles), $= 25$ s (squares) and $= 100$ s (circles).

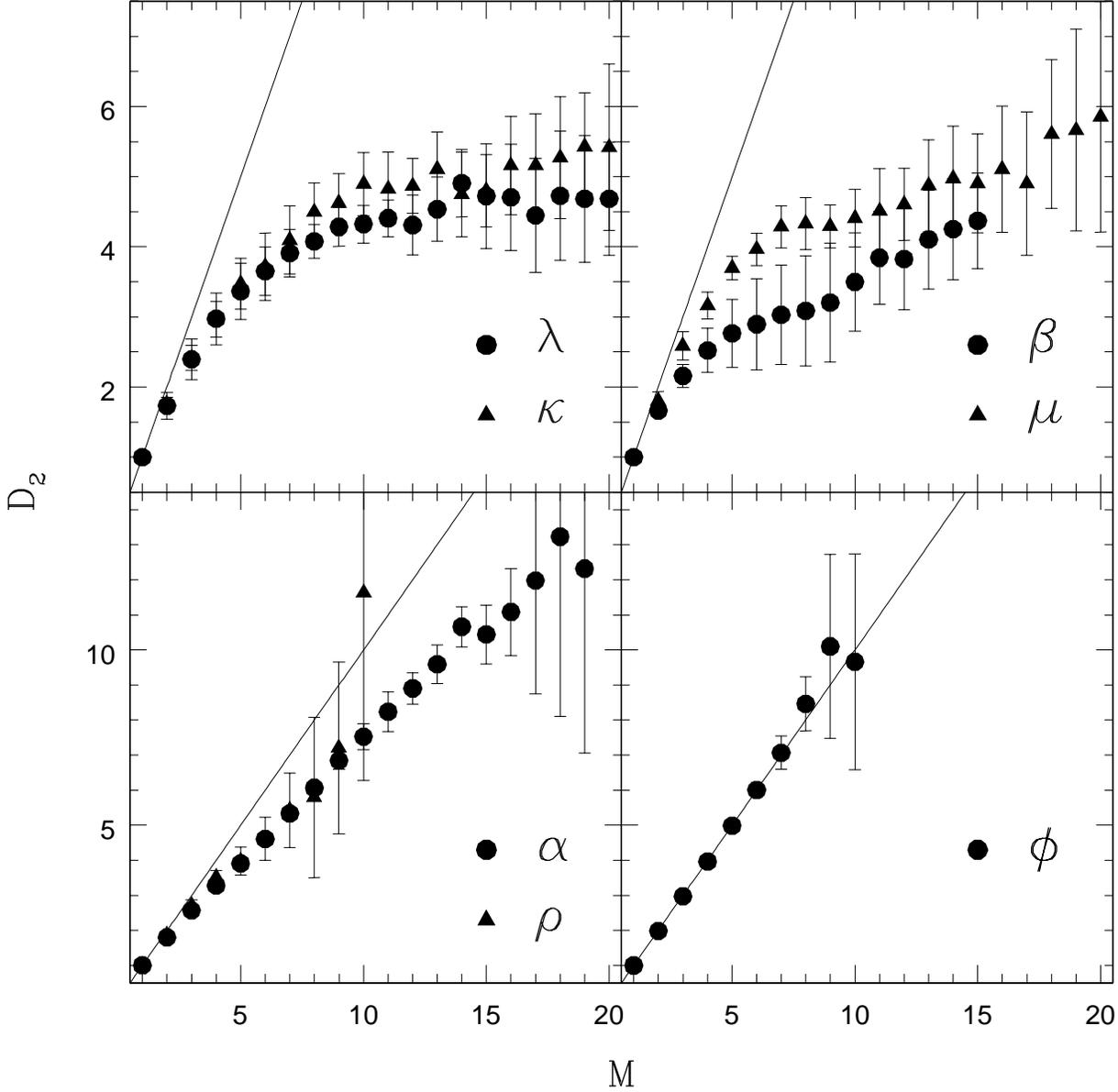

Fig. 2.— The $D_2$ versus $M$ curves for GRS1915+105 data obtained during seven temporal classes. The classes $\kappa, \mu, \beta$ and $\lambda$ exhibit chaotic behavior. The classes $\phi$ is stochastic, while $\alpha$ and $\rho$ show some departure from stochastic behavior.

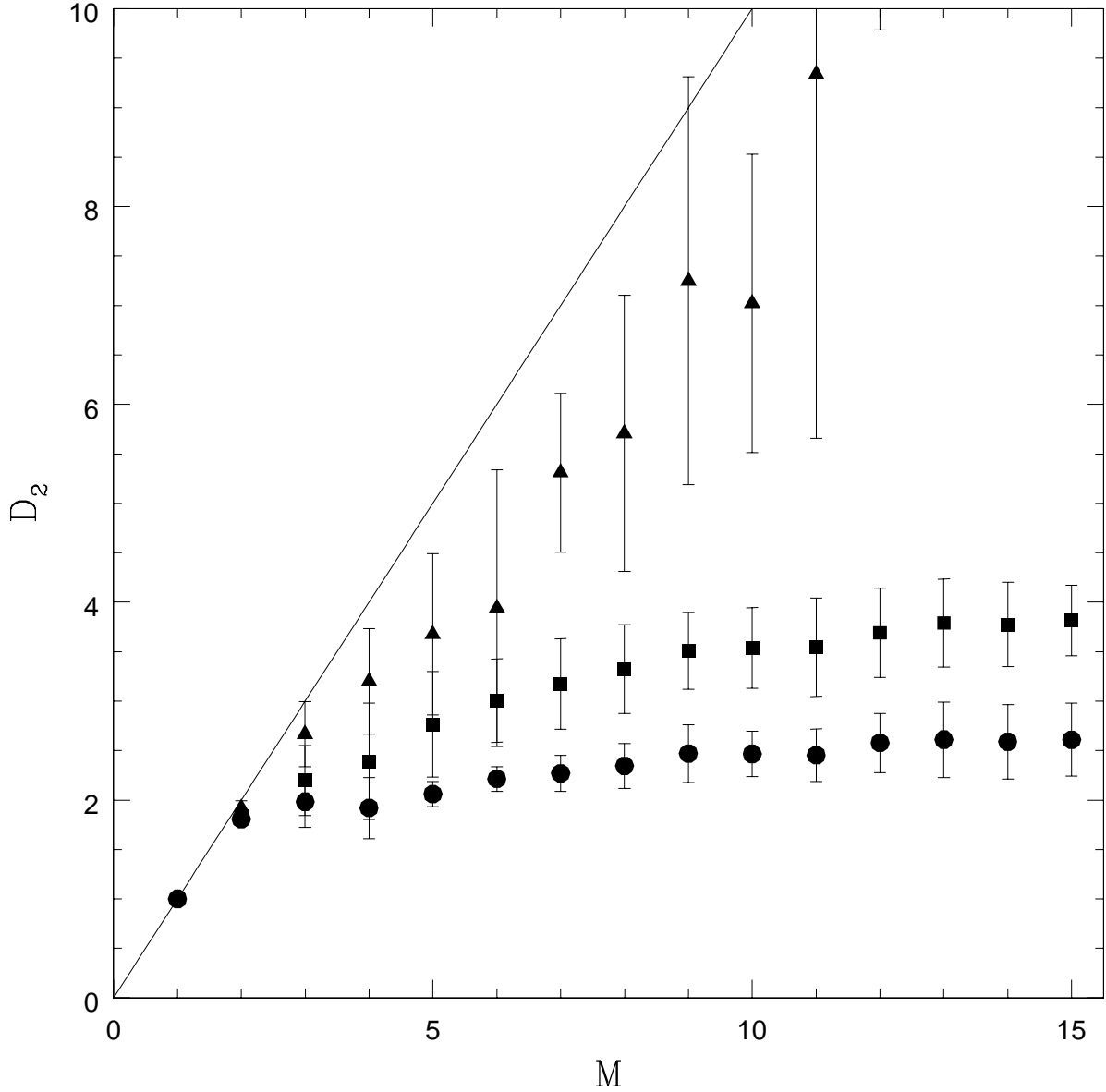

Fig. 3.— The effect of Poisson noise on the analysis is studied using simulated rescaled Lorenz system data with Poisson noise. The $D_2$ versus $M$ curve for Lorenz system is shown by filled circles, while the squares (triangles) are curves for rescaled Lorenz system data corresponding to the expected Poisson noise in the $\beta$ ($\gamma$) class of GRS 1915+105